\documentclass[12pt,preprint]{aastex}
\usepackage{natbib}

\begin{document}

\title{Similar radiation mechanism in gamma-ray bursts and blazars: evidence from two luminosity correlations}

\author{F. Y. Wang$^{1,2}$, S. X. Yi$^{1,2}$ and  Z. G. Dai$^{1,2}$}

\affil{$^1$School of Astronomy and Space Science, Nanjing
University,
Nanjing 210093, China \\
$^2$Key Laboratory of Modern Astronomy and Astrophysics (Nanjing
University), Ministry of Education, Nanjing 210093, China}

\altaffiltext{}{E-mail: fayinwang@nju.edu.cn (FYW); dzg@nju.edu.cn
(ZGD)}

\begin{abstract}
Active galactic nuclei (AGNs) and gamma-ray bursts (GRBs) are
powerful astrophysical events with relativistic jets. In this Letter
the broadband spectral properties are compared between GRBs and the
well-observed blazars. The distribution of GRBs are consistent with
the well-known blazar sequence including the $\nu L_\nu(5\rm
GHz)-\alpha_{\rm RX}$ and $\nu L_\nu(5\rm GHz)-\nu_{\rm peak}$
correlations, where $\alpha_{\rm RX}$ is defined as the broadband
spectral slope in radio-to-X-ray bands, and $\nu_{\rm peak}$ is
defined as the spectral peak frequency. Moreover, GRBs occupy the
low radio luminosity end of these sequences. These two correlations
suggest that GRBs could have a similar radiation process with blazars
both in the prompt emission and afterglow phases, i.e., synchrotron
radiation.

\end{abstract}


\section{Introduction}
Gamma-ray bursts (GRBs) and active galactic nuclei (AGNs) are both
powered by relativistic jets from accreting black holes (Gehrels et
al. 2009; Urry \& Padovani 1995). The central engines of GRBs are
argued to be stellar-mass black holes (Woosley 1993) and for AGNs
the central engines are supermassive black holes. GRBs are most
powerful explosions with isotropic-equivalent energy $E_{\rm
iso}\sim 10^{50}-10^{55}$ erg in the universe (Zhang 2011), and can
be detected out to very high-redshift universe (Lamb \& Reichart
2000; Wang et al. 2012). So GRBs can probe high-redshift universe,
including dark energy (Dai, Liang \& Xu 2004; Schaefer 2007; Wang \&
Dai 2011). Blazars include two subtype of AGNs, i.e., flat-spectrum
radio quasars (FSRQs) and BL Lac objects (BL Lacs). A subclass of
AGNs, e.g., super-Eddington accreting supermassive black holes, are
also proposed to be a standard candle (Wang et al. 2013).

The radiation mechanism of balzars is well constrained. The
spectral energy distribution (SED) of blazars is well understood,
including the low-energy (infrared-soft X-ray) bump and the
high-energy (MeV-GeV) bump. The synchrotron radiation can account for
the low-energy peak, while the MeV-GeV peak is produced by inverse
Compton radiation. But for GRBs, the radiation mechanism for the
prompt emission is still highly debated. The spectrum of prompt
emission can be modeled by the ``Band function'' (Band et al. 1993),
whose origin is still unknown (but see Lucas Uhm \& Zhang 2013).
Some studies (M\'{e}sz\'{a}ros et al. 1994; Daigne \& Mochkovitch
1998) proposed that synchrotron radiation is the leading mechanism.
Other mechanisms are also proposed (Pe'er et al. 2006; Rees \&
M\'{e}sz\'{a}ros 2005; Beloborodov 2010). The radiation mechanism of
afterglows is well understood (Sari et al. 1998). The observed
afterglow radiation is well explained by synchrotron radiation (Sari
et al. 1998; Panaitescu \& Kumar 2001).

Some studies (Zhang 2007; Wang \& Dai 2013; Wang et al. 2014) have
proposed that the mechanisms in different scale outflow or jet
systems may be the same. Some works have been done on comparison
between GRBs and AGNs. Wang \& Wei (2011) compared the spectral
properties of blazars and optically bright GRB afterglows, and found
that GRB afterglows have the same radiation mechanism as BL Lac
objects. A similar correlation of the synchrotron luminosity and
Doppler factor between GRBs and AGNs has been found (Wu et al. 2011).
Nemmen et al. (2012) suggested that the relativistic jets in AGNs
and GRBs have a similar energy dissipation efficiency. Ma et al.
(2014) extended the analysis of Nemmen et al. (2012) by adding X-ray
binaries and low-luminosity AGNS. Wang \& Dai (2013) found that the
GRB X-ray flares and solar X-ray flares have similar distributions,
which indicate that the X-ray flares of GRBs are due to a magnetic
reconnection process. These similar distributions also exist in
X-ray flares from black hole systems with $10^6-10^9M_\odot$ (Wang
et al. 2014). Zhang et al. (2013) found that the prompt emission of
GRBs may be produced by magnetic dominated jets.

In this Letter we compare the broadband spectral properties of GRBs
and blazars, including the $\nu L_\nu(5{\rm GHz})-\alpha_{RX}$ and
$\nu L_\nu(5{\rm GHz})-\nu_{\rm peak}$ correlations, where
$\alpha_{\rm RX}$ is the radio-to-X-ray spectral slope. For a GRB,
$\nu_{\rm peak}$ is the peak frequency of $\nu f_\nu$ spectrum of
prompt emission, while $\nu_{\rm peak}$ is the low peak of $\nu
f_\nu$ spectrum for a blazar. The aim of this Letter is to explore a
possible similarity in radiation mechanism between GRBs and blazars.
this Letter is organized as follows. In section 2,
we present the sample of blazars and GRBs. The fitting results are
given in section 3. Section 4 gives conclusions and discussions.

\section{Samples}
Chandra \& Frail (2012) compiled a sample of GRB radio afterglow
observations from 1997 to 2011. This catalog consists of 304 GRBs
with radio observations. We select 43 GRBs with redshift
measurements. The sample is listed in Table 1. For each GRB, the
name of GRB and redshift are presented in Columns 1 and 2,
respectively. X-ray flux measured in 0.3-10 keV energy band is given
in Column 3 observed by Swift at the time of Column 6. For bursts
observed by BASTE, the energy range is between 2 keV and 10 keV. We
use the typical GRB spectrum to convert the flux to 1 keV. The
observed radio flux (Column 4) and frequency (Column 5) at the time
of Column 6 are also provided. Column 7 is the jet opening angle.
For GRBs without opening angle determination, we assume a typical
value 5 degree. The derived collimation-corrected radio luminosity
at 5 GHz is given in Column 8. Column 9 gives $E_{\rm peak}$, which
is the peak energy of the prompt spectrum. The parameters from
Column 2 to Column 7 are taken from Chandra \& Frail (2012). We use
the value of $E_{\rm peak}$ from Wang, Qi \& Dai (2011). In the
calculation, we use the peak frequency in the GRB rest frame. The
collimation-corrected radio luminosity is calculated as
\begin{equation}
\nu L_\nu(5{\rm GHz})=4\pi d_L^2\nu f_\nu F_{\rm beam}
(1+z)^{-\alpha-1},
\end{equation}
where $d_L$ is the luminosity distance, $f_\nu$ is radio flux at 5
GHz, $F_{\rm beam}=1-\cos\theta_j$ is the beaming factor and
$\alpha=1/3$ is the spectral slope (Sari et al. 1998). We adopt
$\alpha=1/3$ in the slow cooling case. The observed radio flux is
converted to flux at 5 GHz using spectral slope $\alpha=1/3$. In
this work, we assume the cosmological parameters:
$H_0=70$ km s$^{-1}$ Mpc$^{-1}$, $\Omega_m=0.3$, and
$\Omega_\Lambda=0.7$.

We use the spectral properties of balzars from Fossati et al.
(1998). This sample consists of all the parameters that we require,
including redshift, X-ray flux at 1 keV, radio flux at 5 GHz, and
synchrotron peak frequency. Fossati et al. (1998) found a power
spectral sequence for the blazars despite the difference in the
continuum shapes among different sub-classes of blazars. This
sequence indicates that the radio luminosity is anti-correlated with
the synchrotron peak. A plausible interpretation is that
relativistic jets radiate via synchrotron and inverse Compton
processes if the physical parameters (i.e. magnetic field) vary with
luminosity.

\section{Results}
\subsection{$\nu L_\nu(5{\rm GHz})-\nu_{\rm peak}$ correlation in blazars and GRBs}
The radio luminosities at 5 GHz are anti-correlated with the
synchrotron peaks of blazars, as found by Fossati et al.
(1998). This correlation has not been studied in GRBs so far. We
investigate this correlation in GRBs for the first time. Figure 1
shows the $\nu L_\nu(5{\rm GHz})-\nu_{\rm peak}$ correlation of
blazars and GRBs. The black and open dots represent blazars and
GRBs, respectively. There is a tight correlation between $\nu
L_\nu(5GHz)$ and $\nu_{peak}$ as expected from the blazar sequence
(Fossati et al. 1998). The correlation coefficient is $r=0.83$ at a
significance level $p<10^{-4}$ from Spearman rank-order statistical
test. From this figure, we can see that the GRBs occupy the
low-luminosity region of this correlation. For both blazars and
GRBs, the best fitting result is
\begin{equation}
\log \nu L_\nu(\rm 5~GHz)=(-0.91\pm0.03)\log \nu_{peak}+56.65\pm
0.46.
\end{equation}
The correlation coefficient is improved to $r=0.94$ with probability
$p<10^{-4}$. The correlation coefficient has an obvious enhancement
after adding the GRB sample. GRB 060218 may deviate from this
correlation. The possible reason is that this GRB usually called
X-ray flash has a low peak energy (Soderberg et al. 2006; Pian et
al. 2006).

\subsection{$\nu L_\nu(5{\rm GHz})-\alpha_{\rm RX}$ correlation in blazars and GRBs}
The broad-band spectral slope $\alpha_{RX}$ is also correlated with
luminosity at 5 GHz in blazars (Fossati et al. 1998). We also
investigate this correlation in GRBs. The broad-band spectral slope
$\alpha_{RX}$ is defined as
\begin{equation}
\alpha_{RX}=-\frac{\log(f_{\nu_R})/\log(f_{\nu_x})}{\log(\nu_R/\nu_x)},
\end{equation}
where $\nu_R=5$ GHz, and $\nu_X=1$ keV. The X-ray flux $f_{\nu_x}$
at $1$ keV can be obtained as follows. From Column 3 of Table 1,
X-ray flux measured in 0.3-10 keV energy range can be obtained. The
X-ray flux usually evolves as $\nu^{-p/2}$ (Sari et al. 1998), with
$p=2.2$ is the power-law index of accelerated electrons
distribution. The X-ray flux in 0.3-10 keV is $\int_{0.3{\rm
keV}}^{10{\rm keV}}a\nu^{-p/2}d\nu=F_X$. After obtaining the value
of $a$, the flux at $1$ keV can be derived.

Figure 2 shows a correlation between the spectral slope
$\alpha_{RX}$ and radio luminosity at 5 GHz for blazars (black dots)
and GRBs (open dots). The correlation coefficient is $r=0.82$ with
probability $p<10^{-4}$ using Spearman rank-order statistical test
for blazars. After combining GRBs and blazars, the fitting result is
\begin{equation}
\alpha_{\rm RX}=(0.084\pm0.006)\log(\nu L_\nu(5\rm
GHz))+(-2.87\pm0.24).
\end{equation}
The Spearman's rank correlation coefficient is $r=0.85$ with
probability $p<10^{-4}$. GRBs also occupy the the low-luminosity end
of this correlation.

\section{Conclusions and Discussions}
The physics of GRBs are poorly understood, i.e., the radiation
mechanism of prompt emission, the value of Lorentz factor, jet
composition, and central engine (Zhang 2011). Wang \& Dai (2013)
found similar frequency distributions between X-ray flares of GRBs
and solar X-ray flares, which may indicate the magnetically
dominated jets in GRBs. In this Letter we compile 43 GRBs with well
X-ray and radio observations. Two new correlations between
GRBs and blazars may provide a new clue as to the radiation mechanism of
GRB prompt emission and afterglows. For example, our clear $\nu
L_\nu(5{\rm GHz})-\alpha_{\rm RX}$ and $\nu L_\nu(5{\rm
GHz})-\nu_{\rm peak}$ correlations suggest that the radiation
mechanism of GRBs in prompt and afterglow phases and blazars is
similar, namely, synchrotron radiation. Moreover, GRBs are occupy
the low-luminosity region of these correlations.

In this Letter, we use the radio luminosities during the GRB
afterglow phase. Although some models predict that bright radio
emission may be generated within about 10\,s of the initial
explosion of a GRB (Usov \& Katz 2000; Sagiv \& Waxman 2002; Shibata
et al. 2011), but a detection of prompt radio emission has some
impediments, such as scattering (Macquart 2007; Lyubarsky 2008).
Bannister et al. (2012) have searched for prompt radio emission from
nine GRBs at 1.4 GHz, and found single dispersed radio pulses with
significance $>6 \sigma$ in a few minutes following two GRBs.
Unfortunately, the probability of GRB origin is only 2\%. There has
been no confirmed evidence for detection of GRB prompt radio
emission up to now. So we use the radio emission of an afterglow in
this Letter. Theoretically, internal shocks produce the GRB prompt
emission, and external shocks produce the afterglow emission. The
radiation mechanism of an afterglow is well constrained, i.e.,
synchrotron emission. But the prompt emission related to
$\nu_{peak}$ is not well understood. If some similar correlations
between $\nu L_\nu(5{\rm GHz})$ and $\nu_{peak}$ exist in GRBs and
blazars, the radiation mechanism of prompt emission is argued as
synchrotron emission. Until now, the location of blazar gamma-ray
emission regions are still uncertain (Marscher et al. 2010), since
some theories locate blazar gamma-ray emission regions close to the
black hole/accretion disk (Blandford \& Levinson 1995) while the
others place them at parsec scales in the radio jet (Jorstad et al.
2001). The X-ray emission from kiloparsec-scale blazar jets has been
observed (Harris \& Krawczynski 2006). Meanwhile, the radio emission
region of blazars spans kiloparsec scale. Kharb et al. (2010) found
that a few blazars exhibit only radio core emission. But the X-ray
emission region and radio emission region do not fully overlap. So
the high-energy and radio emissions of blazars also originate from
different regions. So a comparison of the correlation $\nu
L_\nu(5{\rm GHz})-\nu_{\rm peak}$ between GRBs and the blazar
sequence is reasonable. Liang et al. (2004) found that the peak
energy $\nu_{\rm peak}$ evolves with isotropic-equivalent luminosity
from the time-resolved spectra. They also found that the
$L_{iso}-\nu_{\rm peak}$ correlation also holds for time-resolved
spectra and time-integrated spectra. Ghirlanda et al. (2010) studied
the time-resolved spectra of Fermi GRBs. The peak energy $\nu_{\rm
peak}$ correlates with the luminosity within individual bursts
(Ghirlanda et al. 2010). Moreover, the time-resolved
$L_{iso}-\nu_{\rm peak}$ correlation is very similar for all the
bursts and has a slope similar to the correlation defined by the
time-integrated spectra of different bursts detected by several
different satellites. The time-integrated value of $\nu_{\rm peak}$
is widely used in luminosity correlations of GRBs, for example
$\nu_{\rm peak}-E_{iso}$ (Amati et al. 2002), $\nu_{\rm
peak}-L_{iso}$ (Yonetoku et al. 2004) and $\nu_{\rm
peak}-E_{\gamma,jet}$ (Ghirlanda et al. 2004) correlations. So if
some correlation is due to a similar physical mechanism, this
correlation holds no matter whether the time-resolved or
time-integrated values are used.

Fossati et al. (1998) found that $\nu_{\rm peak}$ is anti-correlated
with the synchrotron peak luminosity for blazars. For GRBs, Liang et
al. (2004) found that $\nu_{\rm peak}$ is positively correlated with
the isotropic-equivalent luminosity (about total luminosity). But
the luminosity in the correlation for blazars is the synchrotron
peak luminosity, not the total luminosity. Because there are two
peaks in a blazar spectral energy distribution and there is no
correlation between $E_{iso}$ and $\nu L_\nu(5{\rm GHz})$ in GRBs
(Chandra \& Frail 2012), from our simple analysis above we cannot
conclude that GRBs have a different $\nu L_\nu(5{\rm GHz})-\nu_{\rm
peak}$ correlation compared with that for blazars. In this paper, we
find that GRBs occupy the low radio luminosity end of the blazar
sequence, which is similar to that of Wang \& Wei (2011).

\acknowledgements We thank the referee for detailed and very
constructive suggestions that have allowed us to improve our
manuscript. We acknowledge helpful discussions with Y. C. Zou and X.
F. Wu. This work is supported by the National Basic Research Program
of China (973 Program, grant No. 2014CB845800) and the National
Natural Science Foundation of China (grants 11373022, 11103007, and
11033002).

\clearpage
\begin{deluxetable}{lccccccccc}

\tablewidth{0pc} \tabletypesize{\tiny}
\tablecaption{GRB data} \tablenum{1}

\tablehead{
 \colhead{GRB}&\colhead{z}&\colhead{$F_X$}&\colhead{$F_{R}$}&\colhead{$\rm Frequency$}&\colhead{$T_{\rm peak}$}
 &\colhead{$\theta_{j}$}&\colhead{$L_{\rm R}$}&\colhead{$E_{p}$}\\
 \\
\colhead{} & \colhead{} & \colhead{$10^{-13}$erg cm$^{-2}$ s$^{-1}$}
& \colhead{$\mu$JY} & \colhead{GHz} & \colhead{days} & \colhead{}&
\colhead{erg s$^{-1}$} & \colhead{keV} }

\startdata
970508  &   0.835   &   1.2 &   780$\pm$13 &   4.86    &   31.4    &   20.93   &   3.83E39 &   389$\pm$40 &\\
970828  &   0.958   &   8.0    &   144$\pm$31 &   8.46    &   4.0 &   5.15    &   6.70E37 &   298$\pm$30 &\\
980329  &   3   &   ... &   171$\pm$14 &   4.86    &   22.9    &   3.34    &   1.82E38 &   134$\pm$20 &\\
980425  &   0.009   &   3.8 &   38362$\pm$337   &   4.80    &   18.1    &   5.00    &   1.28E36 &   150$\pm$35 &\\
980703  &   0.966   &   4.2    &   1055$\pm$30    &   4.86    &   4.6 &   14.82   &   3.42E39 &   254$\pm$25 &\\
990510  &   1.619   &   2.5    &   255$\pm$34 &   8.46    &   1.6 &   3.36    &   1.27E38 &   126$\pm$10 &\\
010222  &   1.477   &   1.2    &   93$\pm$25  &   8.46    &   6.8 &   3.02    &   3.21E37 &   309$\pm$12 &\\
011121  &   0.362   &   ... &   655$\pm$40 &   8.70    &   5.9 &   40.00   &   2.11E37 &   779$\pm$15 &\\
021004  &   2.33    &   ... &   470$\pm$26 &   4.86    &   5.9 &   13.29   &   5.44E39 &   80$\pm$38  &\\
030226  &   1.986   &   ...    &   171$\pm$23 &   8.46    &   9.7 &   3.43    &   1.24E38 &   97$\pm$17  &\\
030329  &   0.169   &   80   &   10337$\pm$33   &   4.86    &   2.2 &   5.68    &   1.58E38 &   68$\pm$2.2  &\\
031203  &   0.105   &   0.3 &   828$\pm$28 &   4.86    &   28.2    &   5.00    &   3.76E36 &   190$\pm$30 &\\
050416A &   0.65    &   0.47    &   485$\pm$36 &   4.86    &   52.9    &   5.00    &   8.5E37  &   15$\pm$2.7  &\\
050603  &   2.821   &   0.075    &   377$\pm$53 &   8.46    &   29.4    &   5.00    &   9.89E38 &   344$\pm$52 &\\
050820A &   2.615   &   7.7   &   150$\pm$31 &   8.46    &   3.7 &   5.07    &   3.62E38 &   246$\pm$40 &\\
050904  &   6.29    &   0.09 &   76$\pm$14  &   8.46    &   2.6 &   5.00    &   5.69E38 &   435$\pm$90 &\\
051022  &   0.809   &   5   &   268$\pm$32 &   8.46    &   4.8 &   5.00    &   8.57E37 &   510$\pm$20 &\\
060218  &   0.033   &   0.36    &   245$\pm$50 &   4.86    &   2.9 &   80.26   &   2.37E37 &   4.9$\pm$0.3 &\\
070125  &   1.548   &   1.9    &   1028$\pm$16    &   8.46    &   3.7 &   7.90    &   2.63E39 &   366$\pm$51 &\\
071003  &   1.604   &   0.9    &   616$\pm$57 &   8.46    &   33.0    &   5.00    &   6.7E38  &   799$\pm$100 &\\
071010B     &   0.947   &   13    &   227$\pm$114 &   4.86    &   2.5 &   5.00    &   8.12E37 &   52$\pm$6  &\\
080603A     &   1.687   &   0.6    &   207$\pm$26 &   8.46    &   6.4 &   5.00    &   2.45E38 &   85$\pm$30  &\\
090313  &   3.375   &   2.5    &   435$\pm$22 &   8.46    &   1.9 &   3.08    &   5.58E38 &   ... &\\
090323  &   3.57    &   2.0    &   243$\pm$13 &   8.46    &   2.1 &   4.63    &   7.61E38 &   416$\pm$76 &\\
090328  &   0.736   &   2.1    &   686$\pm$26 &   8.46    &   3.4 &   8.85    &   5.74E38 &   592$\pm$237 &\\
090423  &   8.26    &   0.1 &   50$\pm$11  &   8.46    &   9.3 &   9.64    &   1.9E39  &   49$\pm$3.8  &\\
090424  &   0.544   &   2.3 &   236$\pm$37 &   8.46    &   3.6 &   6.33    &   5.62E37 &   162$\pm$3.4 &\\
090715B     &   3   &   1.2 &   191$\pm$36 &   8.46    &   3.4 &   1.39    &   4.23E37 &   134$\pm$40 &\\
090902B     &   1.883   &   4.4    &   84$\pm$16  &   8.46    &   2.3 &   5.00    &   1.19E38 &   775$\pm$11 &\\
091020  &   1.71    &   1.1   &   399$\pm$21 &   8.46    &   4.9 &   5.00    &   4.82E38 &   187$\pm$34 &\\
100414A     &   1.368   &   2.0   &   524$\pm$19 &   8.46    &   4.0 &   5.00    &   4.35E38 &   572$\pm$30 &\\
100418A     &   0.62    &   1.8    &   522$\pm$83 &   4.86    &   3.4 &   8.43    &   2.37E38 &   25$\pm$30 &\\
100814A     &   1.44    &   0.45    &   496$\pm$24 &   4.50    &   43.4    &   5.00    &   3.65E38 &  128$\pm$23 &\\
991216  &   1.02    &   4.6    &   126 &   4.86    &   5.3 &   4.44    &   4.07E37 &   318$\pm$30 &\\
000210  &   0.85    &   ... &   93  &   8.46    &   ... &   5.02    &   3.29E37 &   408$\pm$14 &\\
050401  &   2.898   &   0.8    &   122 &   8.46    &   8.6 &   5.00    &   3.33E38 &   118$\pm$18 &\\
050525A &   0.606   &   0.29    &   164 &   8.46    &   4.6 &   5.00    &   3.01E37 &   81.2$\pm$1.4  &\\
050730  &   3.968   &   1.5    &   212 &   8.46    &   1.5 &   5.00    &   8.92E38 &   104$\pm$20 &\\
050824  &   0.83    &   0.8    &   152 &   8.46    &   8.4 &   5.00    &   5.1E37  &   15$\pm$2  &\\
060418  &   1.49    &   1.5 &   216 &   8.46    &   1.8 &   5.00    &   2.07E38 &   230$\pm$20 &\\
071020  &   2.146   &   1.0    &   141 &   8.46    &   4.0 &   5.00    &   2.45E38 &   322$\pm$65 &\\
100901A     &   1.408   &   7.0    &   331 &   4.50    &   1.7 &   5.72    &   3.07E38 &   ... &\\
100906A     &   1.727   &   2.5    &   215 &   8.46    &   1.0 &   2.88    &   8.77E37 &   ... &\\
\enddata
\end{deluxetable}

\begin{figure}
\begin{center}
\includegraphics[width=\textwidth]{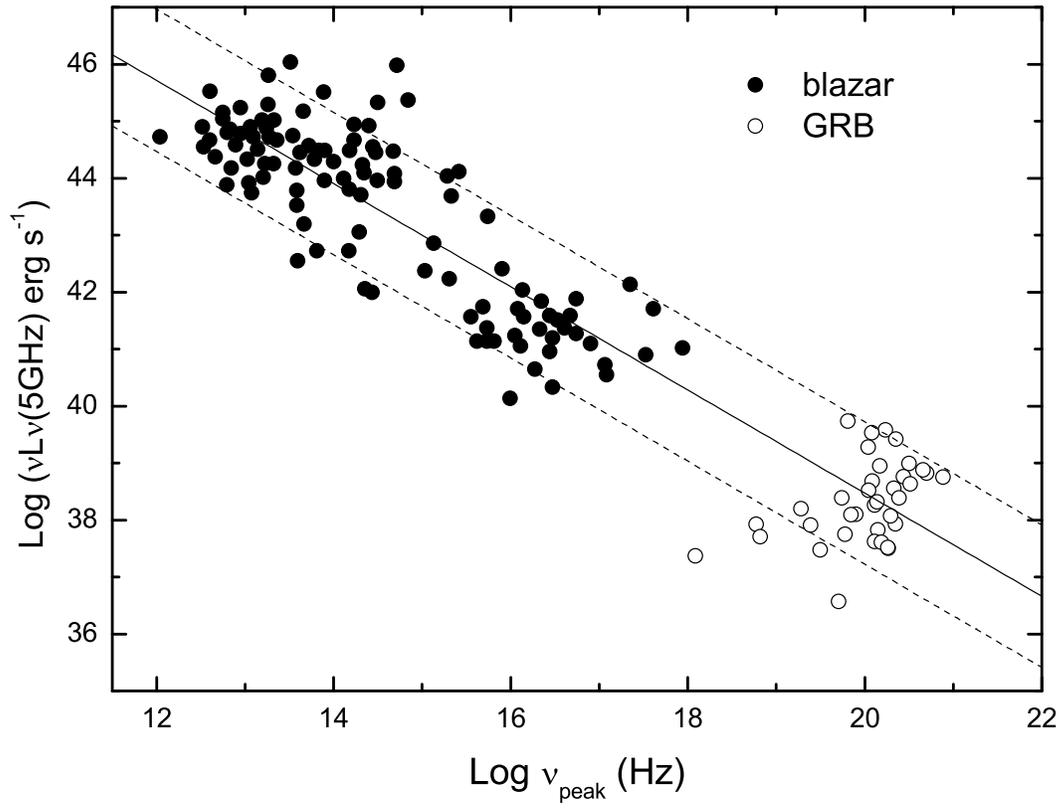}
\end{center}
\caption{Correlation between radio luminosity at 5 GHz and peak
synchrotron frequency for GRBs and blazars. The solid line shows the
best linear fit. The 1$\sigma$ dispersion is marked by the two
dashed lines. \label{fig1}}
\end{figure}

\begin{figure}
\begin{center}
\includegraphics[width=\textwidth]{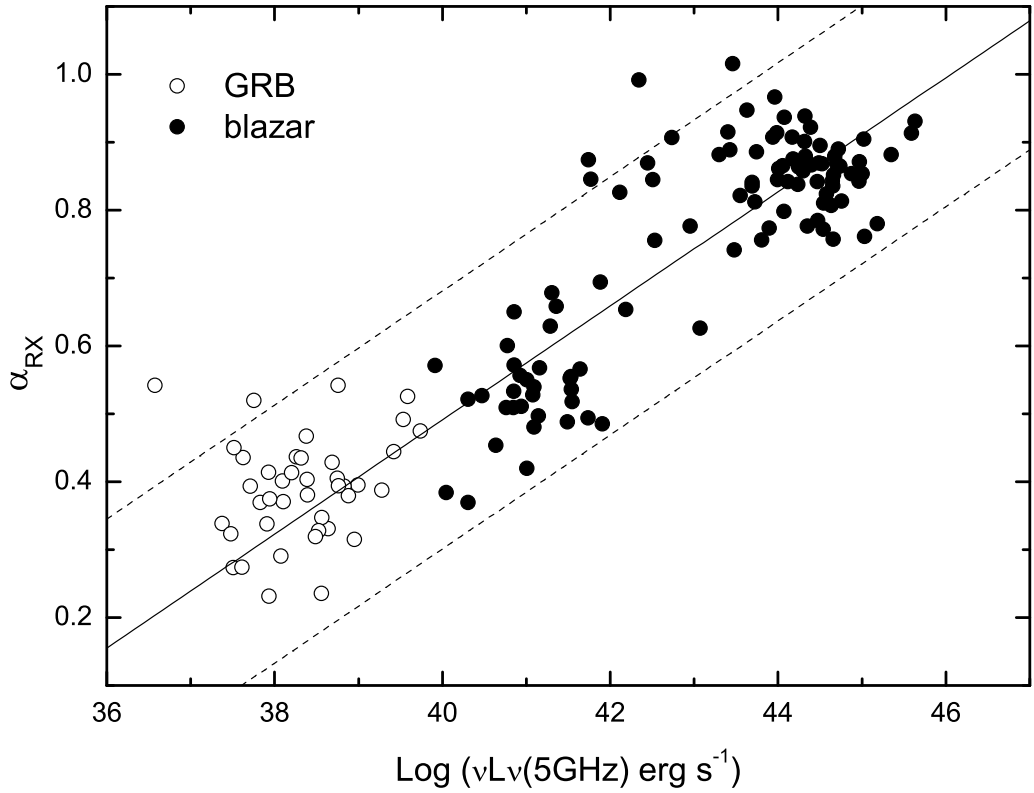}
\end{center}
\caption{Correlation between radio luminosity at 5 GHz and spectral
slope $\alpha_{\rm RX}$ for GRBs and blazars. The solid line shows
the best linear fit. The 1$\sigma$ dispersion is marked by the two
dashed lines. \label{fig2}}
\end{figure}

\end{document}